\PassOptionsToPackage{table, dvipsnames}{xcolor}
 \documentclass[journal]{vgtc}                     %
\onlineid{1328}

\vgtccategory{Research}

\vgtcpapertype{Theoretical \& Empirical}

\title{\hspace{-0.2in}Metrics-Based Evaluation and Comparison of Visualization Notations}

\author{%
  Nicolas Kruchten, Andrew M. McNutt, and Michael J. McGuffin
}

\authorfooter{
  \item
  	Nicolas Kruchten and Michael J. McGuffin are with the École de technologie supérieure.
  	E-mail: nicolas@kruchten.com and michael.mcguffin@etsmtl.ca
  \item
  	Andrew M. McNutt is with the University of Chicago.
  	E-mail: mcnutt@uchicago.edu.

}

\abstract{A visualization notation is a recurring pattern of symbols used to author specifications of visualizations, from data transformation to visual mapping.
Programmatic notations use symbols defined by grammars or domain-specific languages (\eg{} ggplot2, dplyr, Vega-Lite) or libraries (\eg{} Matplotlib, Pandas).  
Designers and prospective users of grammars and libraries often evaluate visualization notations by inspecting galleries of examples. 
While such collections demonstrate usage and expressiveness, their construction and evaluation are usually ad hoc, making comparisons of different notations difficult. 
More rarely, experts analyze notations via usability heuristics, such as the Cognitive Dimensions of Notations framework.
These analyses, akin to structured close readings of text, can reveal design deficiencies, but place a burden on the expert to simultaneously consider many facets of often complex systems. 
To alleviate these issues, we introduce a metrics-based approach to usability evaluation and comparison of notations in which metrics are computed for a gallery of examples across a suite of notations. 
While applicable to any visualization domain, we explore the utility of our approach via a case study considering statistical graphics that explores 40 visualizations across 9 widely used notations.
We facilitate the computation of appropriate metrics and analysis via a new tool called NotaScope. 
We gathered feedback via interviews with authors or maintainers of prominent charting libraries  ($n=6$). 
We find that this approach is a promising way to formalize, externalize, and extend evaluations and comparisons of visualization notations.
}


\teaser{
  \centering
    \includegraphics[width=\linewidth]{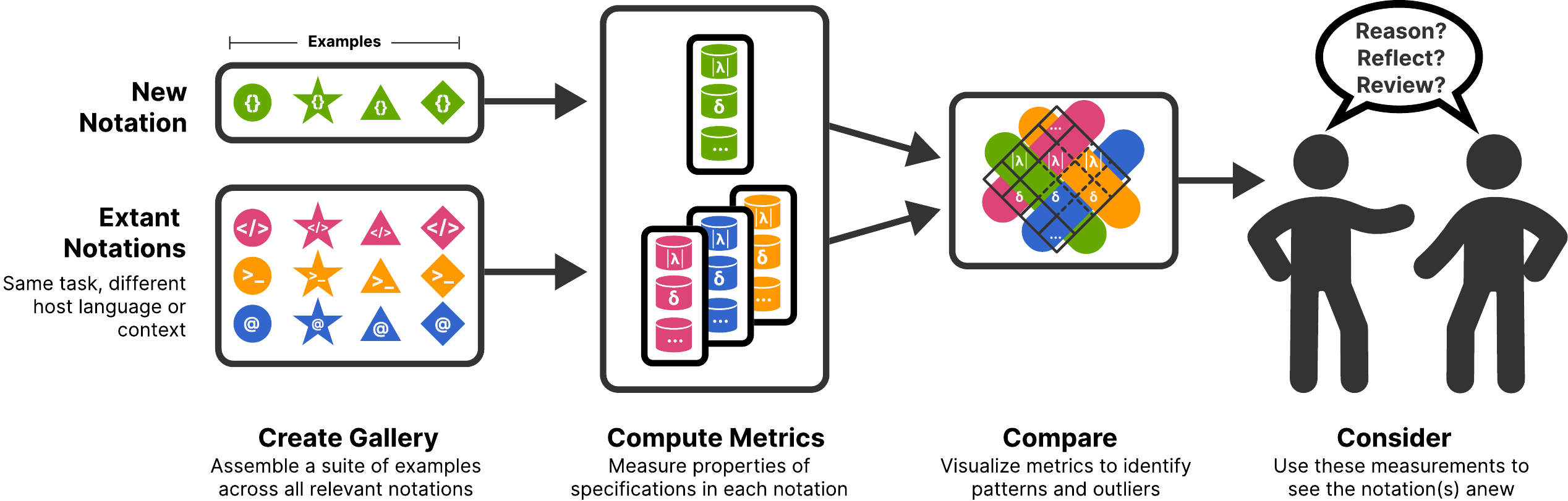}
  \caption{%
  	Evaluating or comparing a visualization notation with others is typically an entirely qualitative, ad hoc process. Our metrics-based approach provides a structured process that automatically produces quantitative data from a set of example specification-visualization pairs to formalize and inform such activities.
  }
  \label{fig:teaser}
}

\graphicspath{{figs/}{figures/}{pictures/}{images/}{./}} %

\usepackage{tabu}                      %
\usepackage{booktabs}                  %
\usepackage{lipsum}                    %
\usepackage{mwe}                       %
\usepackage{soul}
\usepackage{mathptmx}                  %

\newcommand{\etal}
{et al.}
\newcommand{\etals}
{et al.'s}

\newcommand{\ie}{{i.e.}}
\newcommand{\eg}{{e.g.}}

\newcommand{\ala}{\`a la}
\newcommand{\toolName}[1]
{#1}

\newcommand{\metricName}[1]{\textbf{\emph{#1}}}

\newcommand\notation[1]{{\fontfamily{cmss}\selectfont{#1}}}

\newcommand{\vgl}{{\notation{Vega-Lite}}}
\newcommand{\ggp}{{\notation{ggplot2}}}
\newcommand{\px}{{\notation{plotly.express}}}
\newcommand{\go}{{\notation{plotly.go}}}
\newcommand{\sns}{{\notation{seaborn}}}
\newcommand{\sbo}{{\notation{seaborn.objects}}}
\newcommand{\pd}{{\notation{pandas.plot}}}
\newcommand{\mpl}{{\notation{matplotlib}}}
\newcommand{\alt}{{\notation{Altair}}}

\newcommand\dimension[1]{{\color{teal}\emph{#1}}}
\newcommand\code[1]{{\fontfamily{cmtt}\selectfont{#1}}}

\newcommand{\figref}[1]{\hyperref[#1]{Fig.~\ref*{#1}}}
\newcommand{\secref}[1]{\hyperref[#1]{Sec.~\ref*{#1}}}

\newcommand{\footnoteWithIndent}[1]
{\footnote{#1}}

\newcommand{\osf}{\href{https://osf.io/8924y/}{osf.io/8924y}}

\newcommand{\parahead}[1]
{%
  \paraheadd{#1}.
}

\newcommand{\paraheadd}[1]
{%
  \vspace{0.07in}%
  \noindent%
  \textbf{\textit{#1}}%
}

\def\subsubsec#1
{\subsubsection{#1}}

\newcommand{\hlc}[2][yellow]{{%
      \colorlet{foo}{#1}%
      \sethlcolor{foo}\hl{#2}}%
}
\newcommand\qt[1]{\hlc[Periwinkle!15]{``#1''}}

\newcommand{\pxx}[1]{\textbf{E$_{#1}$}}
\newcommand{\pxa}{\pxx{1}}
\newcommand{\pxb}{\pxx{2}}
\newcommand{\pxc}{\pxx{3}}
\newcommand{\pxd}{\pxx{4}}
\newcommand{\pxe}{\pxx{5}}
\newcommand{\pxf}{\pxx{6}}
 
\begin{document}

\firstsection{Introduction}

\maketitle

A \textit{visualization notation} is a recurring pattern of symbols used to author \textit{specifications} of complete visualization pipelines~\cite{chi2000taxonomy}, covering data transformation to visual mapping.
Programmatic notations use symbols defined by grammars or domain-specific languages (\eg{} ggplot2~\cite{wickham2010layered}, dplyr~\cite{dplyr}, Vega-Lite~\cite{satyanarayan2016vegalite}) or libraries (\eg{} Matplotlib~\cite{hunter2007matplotlib}, Pandas\cite{mckinneyProcScipy2010}).
A grammar or library can often be used to achieve a given task in multiple ways, giving rise to multiple alternative notations%
\footnote{We use the term ``notation'' after the Cognitive Dimensions of Notations (CDN) framework~\cite{green1989cognitive} and to signal our focus on the syntactic form and structure of specifications rather than their semantic content.}%
.
For instance, in Python, an analyst might produce an equivalent figure using
Matplotlib or by using higher-level tools such as Seaborn~\cite{waskom2021Seaborn}---the difference between the two being essentially notational.
New visualization grammars or libraries inevitably prompt new usage patterns that give way to new visualization notations and conventions~\cite{bostock2011d3,liu2021atlas, park2017atom, hyeok2022cicero, lee2023deimos, mcnutt2022noGrammar}.
While such innovations provide new functionality and means of expression, evaluating them is a significant challenge~\cite{pu2021special}---despite the variety of available methods.
Empirical evaluations
(such as those that ask novices to recreate extant charts) can measure notational usability or learnability, but their usage can be prohibitively expensive or arduous to execute effectively.
Theoretical evaluation frameworks or discount usability studies~\cite{zuk2006heuristics} (such as the Cognitive Dimensions of Notations~\cite{green1989cognitive} or CDN) are sometimes used to guide close readings of usability~\cite{satyanarayan2016vegalite, bostock2011d3, mcnutt2022noGrammar}.
However, these heuristic methods require considerable skill and can be inordinately subjective.

A prominent~\cite{liu2018data, liu2021atlas, park2017atom, bostock2011d3, li2020gotree, satyanarayan2016vegalite} way to evaluate  notations is through gallery-based expressiveness demonstrations. These consist of a suite of \textit{examples}, where each instance is a (specification, rendered graphic) pair.
These galleries are meant to demonstrate the expressivity of a notation by presenting the breadth and diversity of visualizations it can specify.
However such presentations are often unsystematic in their organization, making more-than-superficial comparisons between notations difficult.
Some practitioners try to facilitate inter-notation comparison through one-off blog posts~\cite{dramaticTour, muth12Libraries} or by developing Rosetta-stone-like galleries~\cite{pythonplot, text2diagram} that consist of a small set of charts expressed in multiple notations.
These usually provide little analysis and few affordances for multi-specification comparisons.
New means of evaluating notations are of high potential value as the audience for programmatic visualization notations is large and impactful---\mpl{} is downloaded $>$35M/month~\cite{pypistats} and is used in $\geq$15\% of arXiv papers~\cite{rocklin}.

\begin{figure}[t]
  \centering
  \includegraphics[width=\linewidth]{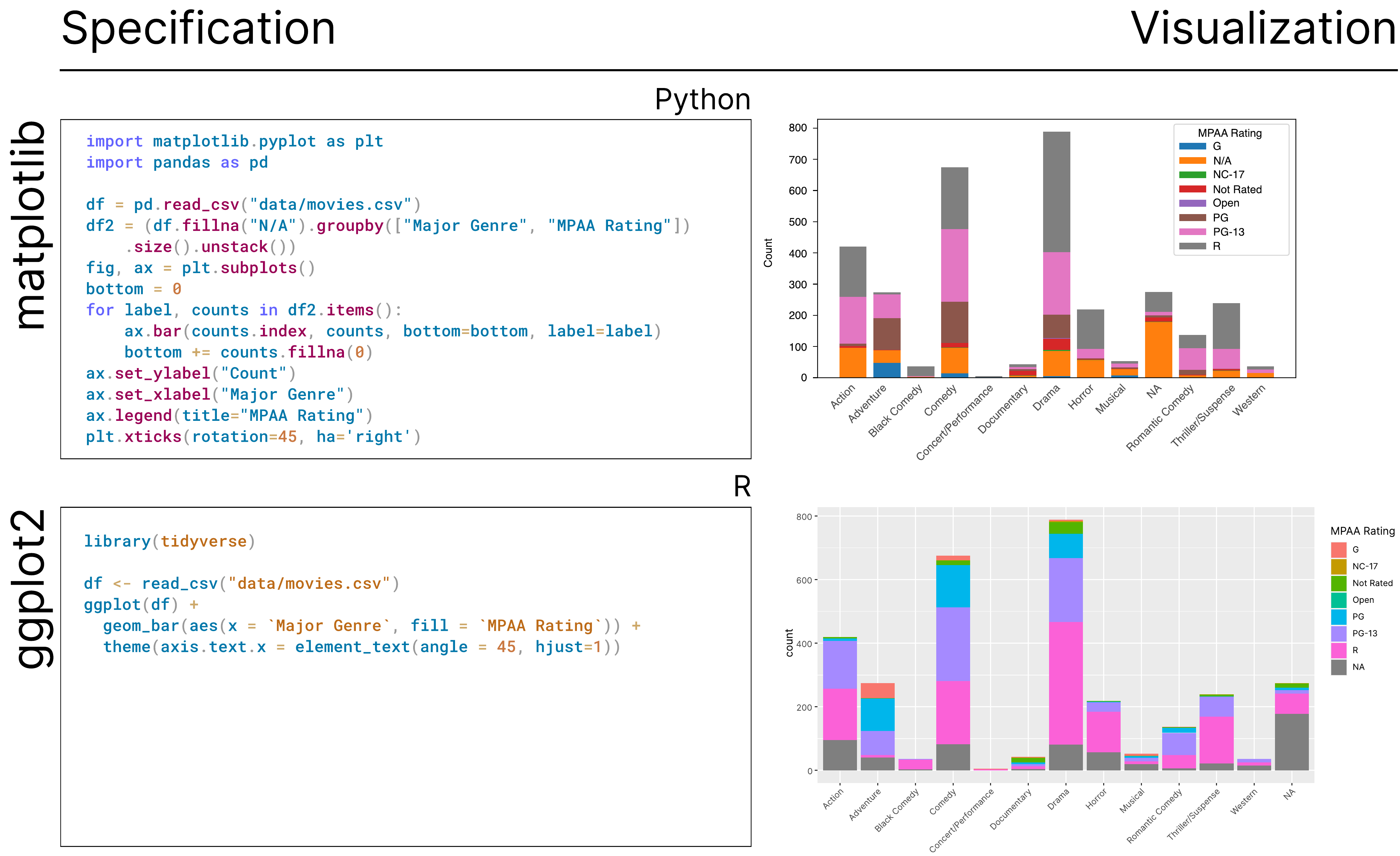}
  \caption{
    Multi-notation galleries consist of a set of notations, each of which implements a shared set of examples or specification and visualization pairs---here a single example of a stacked bar chart.
  }
  \label{fig:specs-example}
  \vspace{-1.5em}
\end{figure}

To address these issues, we introduce a metrics-based approach to evaluating and comparing the usability of visualization notations (\secref{sec:method}).
As outlined in \figref{fig:teaser}, this approach involves forming a gallery of examples (such as from \figref{fig:specs-example}) to create a common point of comparison, computing metrics about that gallery, and then analyzing those measurements as a way to reflect on the properties of a given notation.
These metrics do not carry a normative view of notation design  (\eg{} that they should always be minimized)---rather, they support an externalized way to examine the notations in question.
This approach enables what McNutt \etal{}~\cite{mcnutt2020supporting} would refer to as a \emph{near-by reading} of parallel notations for a common task: a close reading of a familiar text aided by a computational measurement (and hence distanced reading).
We introduce two supporting constructs. First, we propose \emph{multi-notation galleries} that contain a set of examples for a given dataset expressed over a set of notations (\secref{sec:multi-notation-galleries}).
Second, we propose a notion of metrics that can be computed over these structured galleries and provide an initial set of example metrics (\secref{sec:metrics}).
These proposed metrics capture aspects of notational \dimension{viscosity}~\cite{green1989cognitive} (how difficult it is to change specifications), \dimension{economy}~\cite{iverson1979notation} (how many elements and rules a user must keep in mind when using the notation), and \dimension{terseness} (how much can be expressed in a small space), however other notationally-important properties could be metricized as well \footnote{
  We indicate attributes of notations (such as those of CDN) like \dimension{viscosity} and notations themselves like  \mpl{}.
}.

To demonstrate our approach and evaluate its efficacy, we conducted a case study in the context of conventional statistical graphics (\secref{sec:case-study}).
We constructed a gallery covering nine notations (including \ggp{}, \vgl{}, and \mpl{}) across 40 visual forms and computed a variety of metrics as a way to exemplify the types of findings that our approach can yield, for instance, highlighting the tradeoff between notational remoteness and vocabulary size.
While applicable to any domain that supports galleries of specification-and-output example pairs, we focus on static statistical graphics because of its familiarity and the number of similarly-purposed notations.
We presented this study to (n=6) authors or maintainers of the notations studied as a way to elicit feedback on our approach via semi-structured interviews.
We find that it is a promising way to formalize, externalize, and extend expert assessments of notations by providing reproducible examples and comparisons upon which to base judgments and explanations.

Our main contribution is our metrics-driven evaluation method. This is aided by two minor contributions: multi-notation galleries and metrics computable from such galleries.
This work seeks to show the value and viability of our metrics-based approach, rather than to enumerate all useful metrics or to verify the utility of multi-notation galleries outside of evaluation.
We seek to enrich the ways in which notation authors and users  understand and evaluate families of notation designs.

\section{Related Work}
Our work draws upon prior studies that evaluate notations in other domains, and those that focus on visualization specifically.

\subsection{Measuring Notations}

Evaluating programmatic notations requires a measurement method, which may involve theoretically-minded or metric-based approaches.

A variety of different framings for evaluating notations have been developed.
Most prominent among these is the Cognitive Dimensions of Notations (CDN) framework for analytically evaluating usability~\cite{green1989cognitive}. CDN includes 14 dimensions, among them \dimension{viscosity} (how easy it is to make changes to specifications), \dimension{abstraction} (how easy it is to extend the notation), \dimension{closeness of mapping} (how similar the notation is to the target domain), \dimension{progressive evaluation} (how easy it is to check work done to date), and \dimension{hard mental operations}
(how demanding the notation is to working memory).
Similarly, Iverson~\cite{iverson1979notation} laid out five characteristics of good notations, including the ease of \dimension{expression} (informally referred to as expressiveness), ability to \dimension{hide detail} (or \dimension{terseness}), and \dimension{economy} (i.e. small vocabulary).
While providing a useful common language for comparison and analysis,
their definitions make it difficult to employ them in concrete practice (\eg{} with quantitative measurement) or may leave too much open to subjective judgment (so that an inexperienced practitioner might not be able to realize their value).
We draw on these framings to inform our selection of metrics, however other metrics could be formed to support  dimensions not investigated.
The usability of programmatic notations has been studied through the lens of domain-specific languages (DSLs)~\cite{2021UsabilityEvaluationDSL} and libraries' application programmer interfaces (APIs)~\cite{2019SystematicMappingStudy}.
Survey-style instruments have been proposed for quantifying this usability in terms of CDN~\cite{2015QuantifyingUsabilityDomainspecific, 2004MeasuringAPIUsability}, but these are measures of expert judgments and can not be computed automatically.
The API Concepts framework~\cite{2015AutomatedMeasurementAPI} enables automated computation of some metrics, however it is not language-agnostic and
requires a machine-readable API definition as input---an issue that motivates our pragmatics-minded use of metrics over galleries approach.
More generally, the field of empirical software engineering~\cite{shull2007guide, fenton2014software} is replete with metrics for automatically measuring code~\cite{devanbu2016belief}.
A common type of measurement computes the size and relative complexity of a piece of code, such as cyclomatic complexity, which measures the possible  number of execution paths~\cite{fenton2014software}.
Source-code differencing algorithms (such as Gumtree~\cite{frick2018generating}) are also prominent, which compute edit distances between the abstract syntax trees represented (AST) by strings.
We draw on these works in the design and selection of our metrics.
Also related to our work are studies that use metrics-driven methods to help practitioners in other domains select libraries~\cite{de2018library, larios2020selecting}.
However, the metrics used are typically based on metadata (such as popularity) rather than the cognitive usability of those notations.

It has been noted that automatically-computed software metrics often do not directly measure real-world quantities, such as usability or complexity~\cite{fenton2014software}. Our work does not seek to overcome this issue. Instead, we use such measures to augment the close reading of notations that occurs when users interact with galleries of notations.
This strategy is closely related to \emph{near-by reading}~\cite{mcnutt2020supporting}---a form of analysis that draws on the digital humanities notion of \emph{distant reading}~\cite{moretti2013distant} (in which computational measurement is used to replace manual text examination) but situated as an augmentation to the practice of \emph{close reading}~\cite{bares20Close}.

\subsection{Measuring and Evaluating Visualization Notations}

Visualization notations are evaluated in various ways.
Most common among these is to judge the expressive or generative power of a notation with a gallery of examples specified in that notation.
\dimension{Expressive power} is taken to vary with gallery content size and diversity, while \dimension{generative power} is taken to be the presence of novel examples~\cite{liu2018data, liu2021atlas, park2017atom, bostock2011d3, li2020gotree}.
The size and diversity of galleries are sometimes compared directly to discuss relative expressive power~\cite{liu2021atlas, ren2018charticulator}.
While such galleries are useful, they provide limited help to readers seeking to compare the properties of different notations for a given task, as gallery construction is typically informal and varies between notations.
Our work provides a systematic way to construct and analyze galleries across notations. %

Notations have also been analyzed with theory-based heuristics.
CDN has been used to evaluate visualization notations including d3~\cite{bostock2011d3}, protovis~\cite{bostock2009protovis}, Lyra~\cite{satyanarayan2014lyra}, Vega-Lite~\cite{satyanarayan2016vegalite}, and visualization domain-specific languages more generally~\cite{mcnutt2022noGrammar}.
Furthermore, Pu and Kay~\cite{pu2020probabilistic} explicitly linked \dimension{viscosity} to the size of the textual edit distance between specifications, an observation which we draw upon in formulating our remoteness metric.
Closely related to our work, Sarma \etal{}~\cite{sarma2023multiverse} conducted a comparative study of notations for multiverse analysis mediated by CDN.
Whereas they augment this analysis with a lab study, we explore how quantitative metrics can support such analyses.
In each of these studies, the structure provided by CDN is critical as it provides a common language for this style of analysis. However, these dimensions are often applied in an ad hoc manner, leaving the authors to decide which parts are relevant and complicating comparisons across authors.
Other theory-based systematic approaches to analyzing families of visualization have also been explored, such as those based on Algebraic Visualization Design~\cite{pu2020probabilistic, mcnutt2021table}.
While valuable, these forms of analyses have little to say about usability as they can only reason about graphical-semantic properties.
Satyanarayan \etal{}~\cite{satyanarayan2019critical} introduce a critical reflections methodology to precipitate higher-level takeaways from shared experiences developing different sorts of tools---such as in Zong \etals{}~\cite{zong2022animated} analysis of their animation extension to Vega-Lite.
We draw on this approach to evaluate the results of our case study by consulting experts who have experience building and maintaining popular charting libraries.

Measured distances between visualization specifications have been used for various purposes.
For instance, several systems use Vega-Lite specifications as the basis for complex systems of reasoning~\cite{wu2022computableviz, zhao2020chartseer}.
GraphScape~\cite{kim2017graphscape} models the edit-distance between two points in the Vega-Lite design space for the purposes of sequencing visualizations in the presentation of analyses.
Chart Constellations~\cite{xu2018chart} is a tool for navigating the design space spanned by Vega-Lite and models distance as an aggregate of GraphScape's distance, shared data fields in the specification, and keyword tags.
ChartSeer~\cite{zhao2020chartseer} measures distances through an auto-encoder~\cite{kusner2017grammar} trained on Vega-Lite specs. GoTreeScape~\cite{li2022gotreescape} also uses the same encoder to explore distances in the GoTree~\cite{li2020gotree} notation.
Comparisons between the visual manifestation of visualizations have also been explored, such as in semantic chart diffing in notebooks~\cite{wang2022diff} or as the basis of a recommender for responsive visualizations~\cite{kim2021automated}.
Our work is related to these, but focused on analyzing the textual differences between notations as interfaces rather than on using differences to drive recommendations.

\section{Measuring Multi-Notation Galleries}
\label{sec:method}

Our work centers on giving examiners of galleries computational tools to enrich their evaluations and thereby provide a measure of distance to their otherwise close readings of programmatic visualization notations.
We aid this goal by introducing \emph{multi-notation galleries} (\figref{fig:specs-example}), which consists of a set of example specifications (or \emph{specs}) written in notations that might vary between programming languages.
We can then measure aspects of these notations via metrics  that take these galleries as inputs.

Computing metrics over multi-notation galleries (\figref{fig:teaser}) can  surface revealing quantities about the notations under inspection that might not be obvious from closely examining each notation individually---as we explore in our case study \secref{sec:case-study}.
In semiotic terms~\cite{vickers2012understanding}, we are analyzing the representamen (the textual input), rather than the object (the system or its output), or the interpretant (the user's understanding).
While these elements are closely related, they are not identical.
Thus, when we write \vgl{} or \ggp{}, we refer to the notation rather than the implementation.

These measures cannot (and are not intended to) capture the full range of the user experience of the notations under study.
Instead, they are meant to provide an externalized foundation for heuristic evaluations of and critical reflection about notation design.
This approach has potential utility for notation designers (such as in  explaining design choices) and for notation users---who might use this approach to understand how notations relate to each other
(for example, how a new notation compares to a familiar one).

\subsection{Multi-notation visualization galleries}
\label{sec:multi-notation-galleries}

To enable systematic metrics-based comparison of notations, we extend the idea of an example gallery to include multiple notations.
Whereas a typical example gallery is often built to demonstrate the expressiveness of a single notation, we define a multi-notation gallery as one in which each example is expressed in every notation.
The intent of our galleries is different from the opportunistic-programming-informed~\cite{brandt2009two} data exploration of the integrated galleries found in tools like  GALVIS\cite{Shen22Galvis} or Ivy~\cite{mcnutt2021integrated}.
Instead it is to help examiners consider notational differences.

The set of specs should strive to span all common tasks for a given visualization task domain in the same manner that a single notation gallery might try to show all tasks that can be achieved with a given library.
A gallery might be constructed around a task domain such as animation, interaction, graphs, scientific visualization, or, as we consider in our case study (\secref{sec:case-study}), conventional statistical graphics.
Effective examples should each address a given attribute of the domain of interes (for instance, in statistical graphics, how heat maps or histograms are produced) and should be expressed in idiomatic usage of each notation.

We add to this definition the constraint that only a single dataset be used throughout.
This dataset should be typical to the task domain while also supporting as much variation as possible to allow for a maximally large set of examples.
Single-notation galleries (such as those of \vgl{} or \go{}) tend to demonstrate various visual forms or properties of the system by including specs made with multiple (often differently-processed) datasets.
These galleries display both \emph{data variation} and \emph{design variation}.
As we are interested in the notation itself (rather than the data or the documentation value of a gallery), we instead focus on observing  variation within the notation by fixing other factors.

We therefore require that each spec include appropriate data processing,
as notations often interweave their functionality with those of their surroundings.
For instance, \vgl{} includes its own data transformation system (because the host language, JSON, does not include any such tools). In contrast, \mpl{} does not include such elements as it can rely on Pandas for preprocessing.
Including data transforms as
part of the visualization notation aligns with many visualization reference models~\cite{chi2000taxonomy, 2005_GrammarGraphics, mcnutt2020surfacing}.
Further, Pu \etal{}~\cite{pu23Grammar} found that analysts who use a grammar-style~\cite{2005_GrammarGraphics} notation (\ggp{}) do so in a way that is tightly coupled to the use of data-transformation grammars (\notation{dplyr}).
This underscores the need to include all relevant visualization pipeline stages as they capture typical notation usage

Finally, we note that the visualizations for a given example need not be perfectly identical at the pixel level across notations. Rather, we require a \emph{semantic isomorphism} in which the mapping of attributes to properties under study is shared.
Homogenizing the rich default styling that is typically present in visualization systems into a single image type does not capture important parts of how semantic properties are specified---instead it lets the implementation context leak into evaluation of the notation.
For instance, how difficult it is to change the default bar color does have bearing on analyses of data transformation and visual mapping notations for statistical graphics.
\subsection{Gallery Reading Guided by Metrics}
\label{sec:gallery-metrics}

We can measure properties of the notations in our multi-notation galleries by computing the metrics on the specs.
Metrics allow an examiner to ask and answer questions that might be difficult to surface via single-notation galleries alone.
To wit, offering structured ways to contrast with other notations---\emph{how does this notation compare to others?}---as well as means to examine the notation itself---\emph{what are the outliers or unusual cases for this notation?}
The metrics can reveal patterns and outliers which can be investigated via close reading of specs.
We define a metric for a given notation, $N$, as a function taking at least one spec from that notation, $S_{N}$, and returning a number ($\delta_{N}:= (...S_{N}) \Rightarrow Number$).
To be useful, $\delta_N$  should credibly correlate with some notational property of interest---such as the CDN properties as we explore in \secref{sec:metrics}.
The measurements can be used either individually (such as our \metricName{vocabulary size}, \secref{sec:vocab}) or in aggregate (such as our \metricName{median specification length}, \secref{sec:terseness}).
This intentionally broad definition allow us to relate specs within a notation as well as to characterize the differences in distributions between notations.

Between notations, we can compare the distributions of metrics to contrast various design strategies.
For instance, pairs of notations can be compared by plotting the aggregate measures against each other and looking for broad patterns or outliers as in \figref{fig:matplotlib_remoteness} and \figref{fig:vega-lite_remoteness}.
Patterns among outliers can be understood by looking for common notation fragments among the underlying specs, so as to draw high-level conclusions about the similarities and differences between notations.
We demonstrate such analyses in our case study in \secref{sec:analysis}.

\begin{figure}[t]
  \centering
  \includegraphics[width=\linewidth]{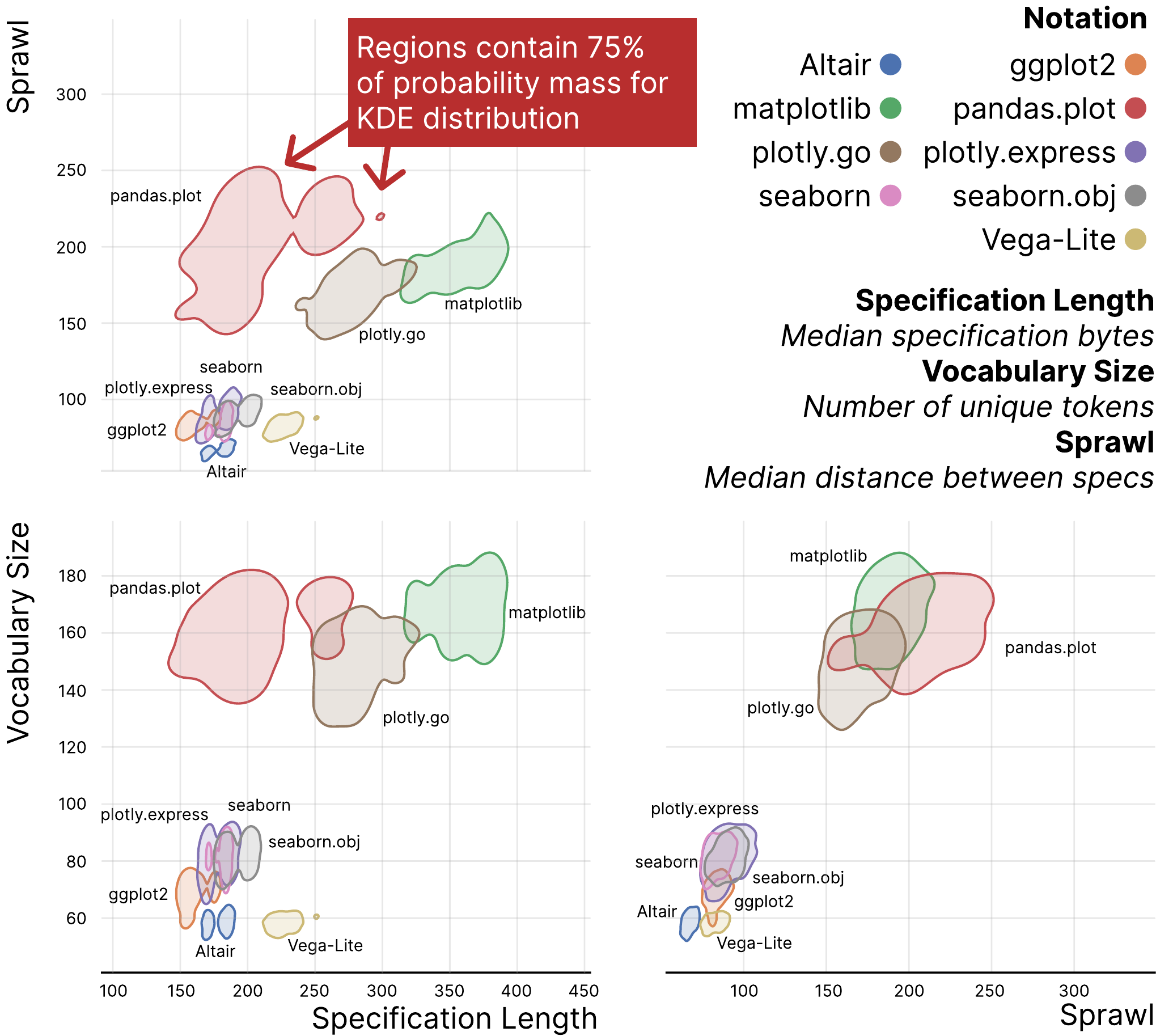}
  \caption{
    To mitigate the effects of factors like outliers or experimenter bias,
    we computed 1k bootstrapped variations for each metric in our case study, which we show here via a Kernel Density Estimation (KDE). The size, proximity, and overlap between regions supports reasoning about the uncertainty inherent in metrics-based analyses of galleries.
  }
  \label{fig:clusters}
  \vspace{-1em}
\end{figure}

For comparisons of more than two notations, plotting the aggregated metrics (such as \metricName{vocabulary size}) against each other can reveal high-level patterns.
However, as many notations support the creation of a given chart in multiple equivalent ways, constructing univariate aggregates may be hallucinatory~\cite{mcnutt2020surfacing} as the gallery may be biased by an analyst's coding style or preferred idioms.
The effect of such biases can be identified by quantifying uncertainty latent to aggregate measures, such as by having a second analyst reconstruct part of the gallery (which would support a $\delta_{error}$ term that could be propagated through any aggregates) or via statistical procedures (such as bootstrapping within $\delta_N$'s arguments).
In \figref{fig:clusters} (and our case study), we express the uncertainty in our gallery by bootstrapping the arguments for several metrics $\delta_N$ (within each notation).
This allows us to explore the relationship between notations via aggregate measures without potentially self-deceiving (solely) based on our construction methods.

Within a notation, metrics that accept pairs of specs can be used to form a distance matrix, which can be used to inform subsequent analyses.
For instance, we identify which specs are closest to each other in the metric space---such as highlighting the connection between pie charts and univariate stacked bar charts in grammar-style notations (\eg{} \ggp{}).
We can use the columns of this matrix as the basis for analyses, such as via dimensionality reduction (\eg{} MDS~\cite{mead1992review} or UMAP~\cite{mcinnes2018umap}) or agglomerative clustering (\eg{} dendrograms) as in \figref{fig:seaborn-obj}.
The study of these patterns of distance can give us insight into how the notation structures the visualization design space spanned by the gallery.

\subsection{NotaScope}

To facilitate this analysis process and to support our case study (\secref{sec:case-study}), we developed a web-based tool called NotaScope.
It provides modular and extensible support for multi-notation galleries whose notations may range across Python, R, JavaScript (JS), and JSON---with parsing and tokenizing provided by the language-agnostic Tree Sitter~\cite{treesitter}.
It processes specs and computes an extensible set of metrics.
It standardizes textual representations of the galleries (to reduce the effect of coding style on computed metrics)  through standard reformatters such as jq~\cite{jq} for JSON, black~\cite{black} for Python, and styler~\cite{styler} and formattr~\cite{formatr} for R.
The web-based front-end supports multiple visualization modes for comparing pairs of specs, visualizing single- and multi-notation galleries, as well as the relationship between notation distance matrices.
See supplement for a demo, source code, and video walk-through.

\section{Notation Metrics}
\label{sec:metrics}

We now describe a set of example metrics that measure various notationally-important quantities from multi-notation galleries.
We introduce metrics that measure \dimension{terseness}~\cite{green1989cognitive}, \dimension{economy}~\cite{iverson1979notation}, and \dimension{viscosity}~\cite{green1989cognitive}.
These dimensions are of interest to us as they characterize prominent aspects of the experience of using these notations to author visualizations, and allow us to make structured comparisons between specifications and notations more generally.
Crucially, these metrics do not encode normative preferences but rather provide their users with conceptual scaffolding on which to form analyses of notations.
We stress that our intent is to demonstrate the feasibility and usefulness of the metrics-based approach, rather than provide a comprehensive accounting of all useful metrics.

\subsection{Terseness: Specification Length}
\label{sec:terseness}

We begin by defining a measure of \dimension{terseness}~\cite{green1989cognitive} that considers the aggregate \metricName{specification length}.
This measure enables analysis of the volume of symbols required to produce a result in different notations.

A natural first measure of any program's source code is its length.
In empirical software engineering, measures of length (such as source lines of code or SLOC) are often used as a crude measure of complexity~\cite{fenton2014software} (as longer code is more complex to modify).
While imperfect~\cite{texel2015measuring}, it is able to highlight the variation in syntactic \dimension{terseness}.
A variety of related measures---such as the cyclomatic complexity or AST-based measures (\eg\ breadth or depth)~\cite{fenton2014software}---are readily available.
Unfortunately, these do not fit our context as we study the variance between  different languages.
For instance, comparing \vgl{}'s nesting JSON-style with \ggp{}'s additive R-style structure will likely only reveal properties related to the conventions of their host languages, rather than the notations.

Instead, we define the \metricName{specification length} of a specification $S$ as the size of $S$ in bytes of UTF-8-encoded text. This circumvents common issues with SLOC~\cite{texel2015measuring}, such as the definition of a line.
Within a notation, per-specification lengths capture a measure of the relative specification verbosity in a way that is not wholly tied to the context of the host language.
We use the \metricName{median specification length} for a notation $N$ across a gallery as a noisy-but-consistent measure of $N$'s \dimension{terseness}.
As discussed in \secref{sec:interviews}, experts reviewing our case study seemed to use size as a proxy for complexity.

\subsection{Economy: Vocabulary Size}
\label{sec:vocab}

Next, we examine notational \dimension{economy}~\cite{iverson1979notation} through the lens of \metricName{vocabulary size}. This measure allows us to consider the tensions between adding terms and grammatical rules in a notation.

When adding new functionality to a notation, designers must either add new ways to combine existing terms or to add new terms to the vocabulary.
For instance, a visualization notation might introduce low-level primitives that can combine to create particular chart types (such as in \ggp{}), whereas another might introduce terms to denote those specific chart types (such as in \px{}). Both approaches typically add new tokens, such as keywords, operators, or functions.
Yet, in their limits, neither option is ideal as they would yield \dimension{hard mental operations}: a huge vocabulary would tax memory, and keeping track of the interactions between many similar tokens would lead to a high cognitive load.
This highlights a tradeoff between \dimension{terseness} and the vocabulary size associated with a notation, which is akin to  Iverson's~\cite{iverson1979notation}'s notational \dimension{economy}.

To explore these issues,
we define the \metricName{vocabulary size} of a notation as the number of unique tokens it uses to specify the examples in the gallery---analogous to Halstead's measure of vocabulary~\cite{fenton2014software}.
We measure this quantity by counting the number of unique tokens across all specs in a gallery---which, in the case of our prototype, is done via the language-agnostic Tree Sitter~\cite{treesitter}.
While most galleries do not enumerate the entire breadth of a notational design space, we can assume that they capture a representative sample as they are meant to familiarize users with the scope of the available functionality.
This metric captures both the effects of the design of the libraries used in the notation and those of any host language, such as Python or JSON, by counting tokens such as \code{import} and punctuation marks.

This metric has limitations: it does not directly measure the number of high-level concepts that a user must learn to use a notation (the actual cognitive burden to address in notational economy), but it is an easy-to-compute approximation of this quantity.
A more complete way to measure the vocabulary size of a notation would be to identify all the individual tokens it can admit through source-code analysis of the underlying system.
This can be prohibitively complicated as although some notations (\eg\ \go{} and \vgl{}) include explicit representations of their language (as JSON Schemas~\cite{pezoa_foundations_2016}), the majority do not have such a computationally tractable API definition---a gap which informs our use of metrics over examples. Further, notations are often cultural conventions that exist outside of the formal definition of a language, and so such attempts may not capture important aspects of usage---such as the un-required but common connection between \mpl{} and Pandas.

\subsection{Viscosity: Distance, Remoteness, and Sprawl}
\label{sec:viscosity}

Finally, we consider notational \dimension{viscosity}~\cite{green1989cognitive},
or the cost of changing one spec into another,
which we define as the \metricName{distance} between specs.
This supports measurements of how much a spec would need to change to become another and allows us to analyze the size of the notation through properties like \metricName{remoteness} and \metricName{sprawl}.
A common first choice for the distance between programs (and strings more generally) is the Levenshtein edit distance (LD), which measures the number of changes required to make two strings equal.
However, LD has several drawbacks for our use case.
LD is high for semantically small changes, for instance changing \code{f(x=a, y=b)} into \code{f(y=b, x=a)} yields a large edit distance for a meaningless rearrangement in Python.
Furthermore, it is very sensitive to the tokenization process, as every pair of tokens is considered equally different.
This fails to capture patterns that are easily recognized by humans:
for example, the \ggp{} tokens \code{geom\_point} and \code{geom\_line} are more similar to each other than \code{geom\_point} and \code{facet\_wrap}.
Similarly, humans' tendency to chunk information leads us to perceive a sequence of 6 tokens such as \sbo{}'s \code{so.Agg("mean")} as quite similar in size and complexity to the 3 tokens of \px{}'s \code{histfunc="mean"}, making cross-notation comparisons of LD values more challenging.
Finally, certain notations have complex tokenization rules: \vgl{} cannot be completely tokenized at a JSON level, as some strings are JS fragments.
Related issues arise in any token or edit distance-based measures, including those that operate on ASTs~\cite{frick2018generating}.

To address these issues, we draw on Li \etals{}~\cite{li2004similarity} ``universal similarity metric'' from algorithmic information theory.
This metric takes into account all possible computable similarities between strings, without requiring any assumptions about the data or operations.
They introduce a practical version of their idealized metric by using a standard compression algorithm (\eg\ gzip or lzma) to separately compress two strings, $A$ and $B$, as well as their concatenation, $AB$, which they refer to as Compression Distance. They define it as
\vspace{-0.5em}
$$
  CD(A,B) = C(AB) - \min( C(A), C(B) )
$$
where $C(X)$ is the compressed length in bytes of $X$.
While it is more challenging to interpret, CD succeeds where LD falls short by  capturing intra- and inter-token regularities and requiring no tokenization (allowing it to be easily applied to any textual notation).
For example, $LD$\code{("geom\_point", "geom\_line")} = $LD$\code{("geom\_point", "facet\_wrap")} = 1 for these single tokens.
In contrast, with zlib as a compressor, $CD$\code{("geom\_point", "geom\_line")} = 7 which is less than $CD$\code{("geom\_point", "facet\_wrap")} = 10, as intuition would suggest.
We normalize the inputs to this computation as a preprocessing step by applying standard code formatting tools (\eg{} code prettification and key sorting).
Beyond being more expressive and less sensitive to non-meaningful aspects of the specs, CD is able to act as a drop-in replacement for LD in our case, as they were closely correlated for distances in our case study gallery (\secref{sec:case-study}), with a Pearson's R of 0.977 ($p<0.001$).
Examination of outliers led us to believe CD is a better measure of  \dimension{viscosity}.
Thus, we define the \metricName{distance}  in notation $N$ between specs $S_{N1}$ and $S_{N2}$ to be $CD(S_{N1}, S_{N2})$.

Given this distance metric, we can describe several other notational properties.
We define the \metricName{sprawl} of a notation $N$ across a gallery as the median distance between all pairs of specifications in $N$.
Next, we define the \metricName{remoteness} of a specification $S_N$ within a gallery as the median distance between $S_N$ and every other specification in the same notation.
Specs with comparatively low remoteness relative to sprawl are considered central to the gallery,
as they require fewer changes to become other specs.
Conversely, those with relatively high remoteness are comparatively hard to get to from the others, and so may be outliers.

\subsection{Further Metrics}
\label{sec:more-metrics}

Beyond the metrics described above, various others might usefully be considered to capture other notational qualities of interest. A useful metric might measure the distance between domain effect and textual control or \dimension{closeness of mapping}~\cite{green1989cognitive} in CDN terms.
Further metricizing the rest of CDN~\cite{green1989cognitive} or Iverson's heuristics~\cite{iverson1979notation} would complement these efforts.
Metrics that are sensitive to programmatic usage context (such as extensibility or maintainability) would be useful.
McNutt~\cite{mcnutt2022noGrammar} argued for a CDN-style evaluation mechanism tuned to the specifics of visualization domain-specific language design. Such a suite of heuristics, if metricized, would be a natural point of comparison between notations.
Other domain-specific metrics would also be useful (such as expressivity of elements like legends, annotations~\cite{mcnutt2022noGrammar}, or accessibility features).
We note that this approach is not bound to statistical graphics and could be applied to other domains (\eg{} graphs, maps, or diagrams).

\begin{figure}[t]
  \centering
  \includegraphics[width=\linewidth]{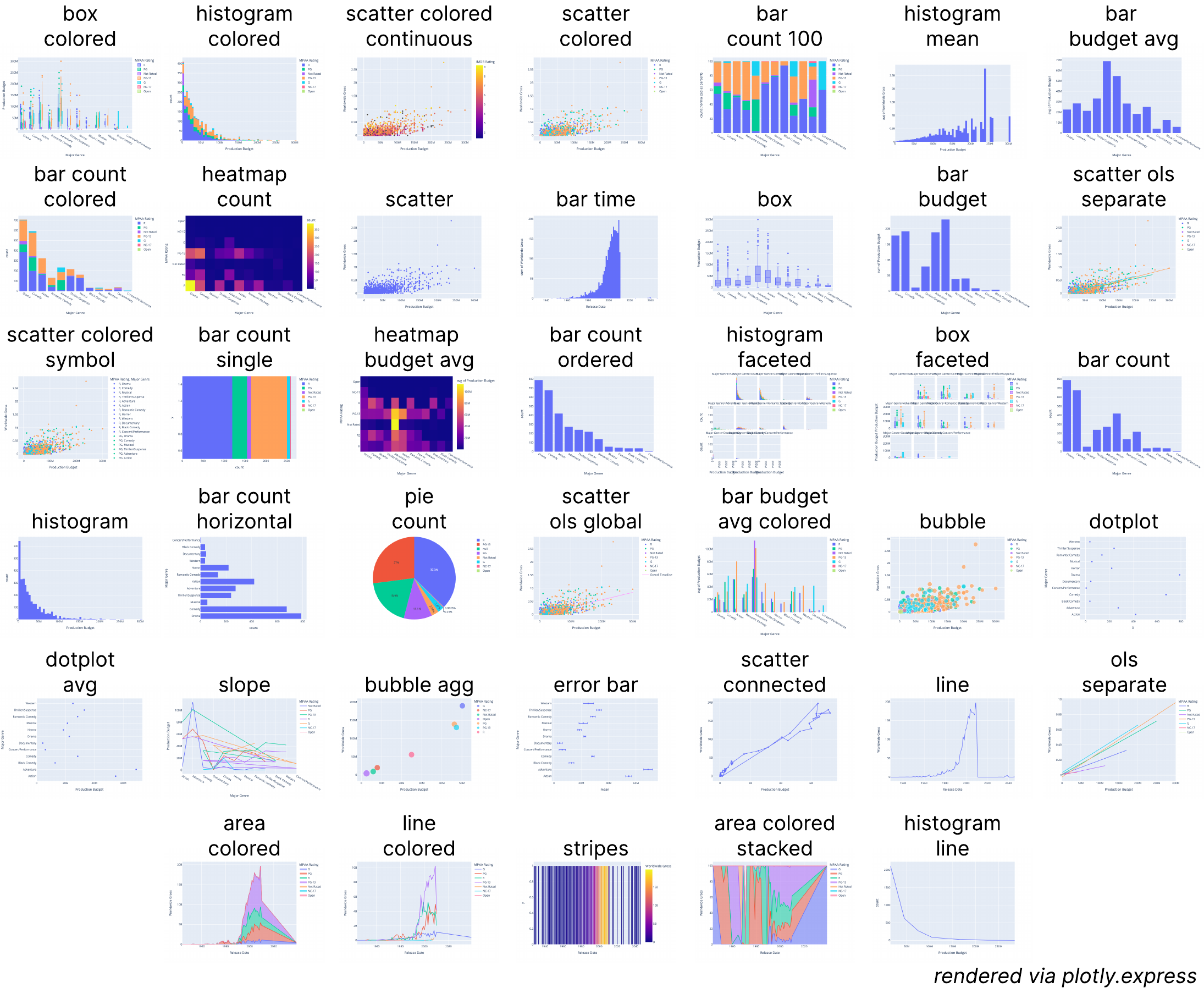}
  \caption{
    The 40 examples used in our case study gallery. In selecting this set we sought charts that span common statistical graphics tasks.
  }
  \label{fig:gallery}
  \vspace{-2em}
\end{figure}

\section{Demonstration: a Statistical Graphics Gallery}
\label{sec:case-study}

Having described our approach and practical metrics to drive it, we now present a case study to demonstrate the application of these tools to evaluate and compare popular notations used for specifying statistical graphics.
We begin by describing an example multi-notation visualization gallery covering this task domain, on which we compute relevant metrics using NotaScope.
We provide an example of the style of analysis in \secref{sec:analysis}.
To evaluate our analyses and approach, we interviewed authors or maintainers of the systems we considered in \secref{sec:interviews}.

\parahead{Notations}
Our gallery covers nine notations:
\ggp{} in R, \vgl{} in JSON and Python (as \alt{}),
as well as two variations each of Matplotlib (\mpl{} and \pd{}), Seaborn (\sbo{} and \sns{}),
and Plotly (\px{} and \go{}) in Python.

We focused on these notations as they are widely used among data science practitioners~\cite{Mooney22Kaggle}, have differing relationships with data and graphic, and are commonly used for typical statistical statistical graphics tasks~\cite{russell2016simple}.
We exclude notations focused on other information visualization tasks (\eg\ building interactive or animated visualizations, visualizing networks, making maps or tables, or rendering three-dimensional data) as they possess idioms which are disjoint from those of basic charting---although similar analyses could be usefully conducted to identify and understand those notations using different galleries.
We include the \pd{} visualization notation as distinct from \mpl{} due to the popularity of pandas and to study the effects of a minor variant of a notation.
Similarly, we include the recent \sbo{} (a new, composable alternative API within the Seaborn library) to study the differences between it the more established \sns{} notation, as they share many design decisions.
Beyond these criteria, we were guided by notations for which we were able to solicit expert feedback to evaluate our analyses, which we describe in \secref{sec:interviews}.
We further describe these notations and detail our inclusion criteria in the appendix.

\begin{figure}[t!]
  \centering
  \includegraphics[width=\linewidth]{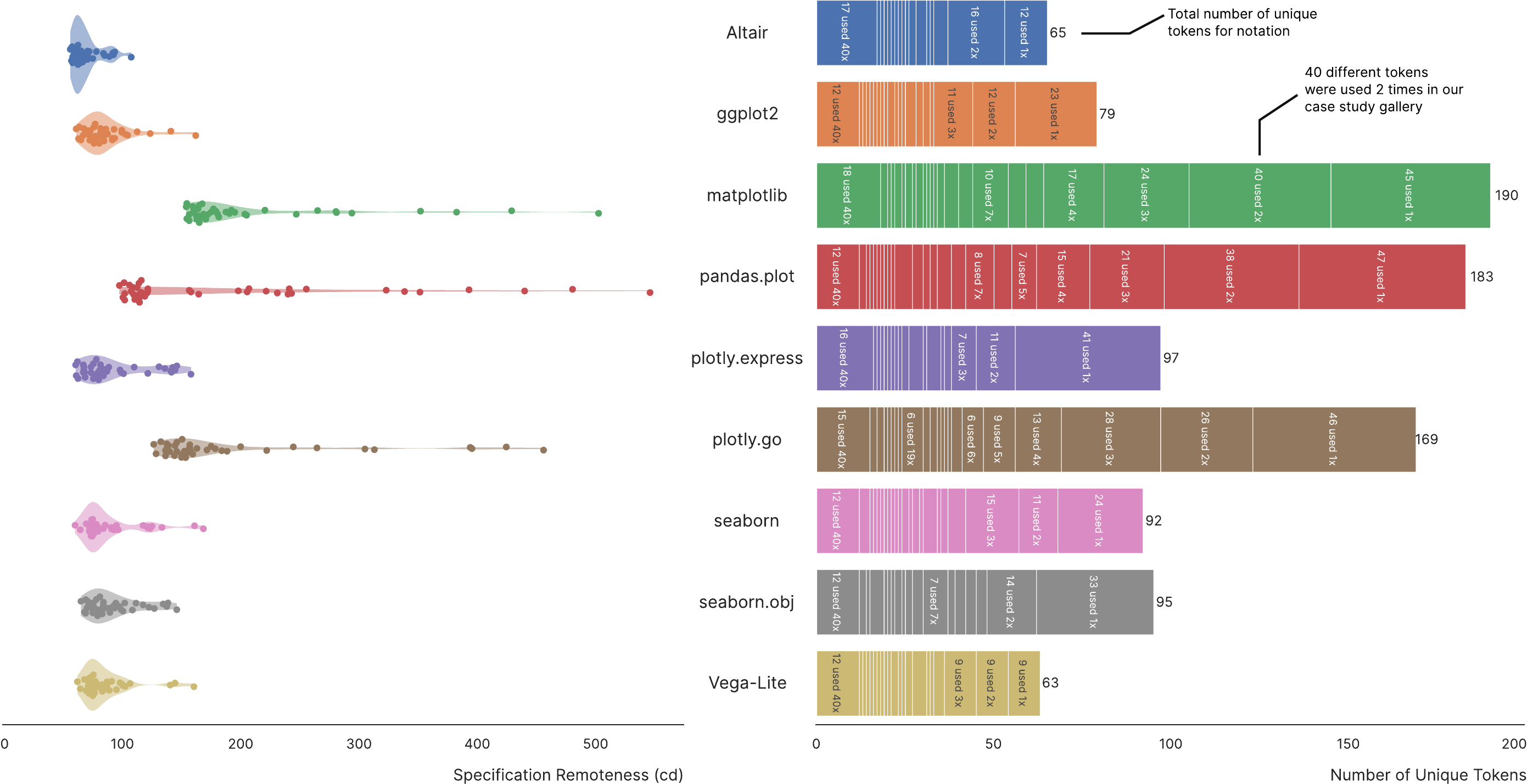}
  \caption{
    Left: The per-notation distribution of remoteness shows a clear ``core and tail'' pattern. Right: The decomposition of unique tokens per notation, broken down by how often the tokens are used in the gallery, suggests a direct relationship between vocabulary size and sprawl.
  }
  \label{fig:remote-distro}
  \vspace{-1em}
\end{figure}

\parahead{Examples and Dataset}
To study these notations, we built a suite of 40 examples (\figref{fig:gallery}) based on the frequently-used tidy~\cite{tidyData} \texttt{movies} dataset~\cite{vegaDatasets}.
We focused on this dataset because it has good coverage of variable types---including continuous, ordinal, and temporal variables---allowing us to create a wide variety of chart forms.
These visualizations cover a variety of conventional chart types (\eg{} bar charts, line charts, scatter plots, and heatmaps) as well as tasks (\eg{} distributions of variables and the relationships between them).
Moreover, they demonstrate a representative sample of statistical graphics tasks and techniques, ranging across mapping of continuous and categorical variables to spatial or color axes, the use of grouping, stacking, binning, aggregation, small multiples, regression, and error bars.
A subset of expert interviewees (\secref{sec:interviews}) iteratively reviewed and guided the curation of this list as a way to achieve ecological validity.

We strove to create idiomatic versions of each chart---such as a typical spec author might write given the constraints of a typical environment---by following the  documentation wherever possible and falling back to strategies espoused in well-rated StackOverflow posts otherwise.
We describe our per-notation construction process in the appendix.
We further seek to limit these biases by visualizing the uncertainty in our  metric values, as in \figref{fig:clusters}.

\subsection{Analysis}
\label{sec:analysis}

We now demonstrate a metrics-based analysis of a multi-notation gallery.
We do so by exemplifying this style of analysis by considering the question \emph{what is typical for notations in this context?} at a series of granularities starting from high-level relationships and zooming into a single notation.
At each level of analysis, we identify patterns that we confirm and explain via spec-by-spec analysis to gain insight into these notations---thereby performing a \emph{near-by reading} of the gallery.

\begin{figure}[t!]
  \centering
  \includegraphics[width=\columnwidth]{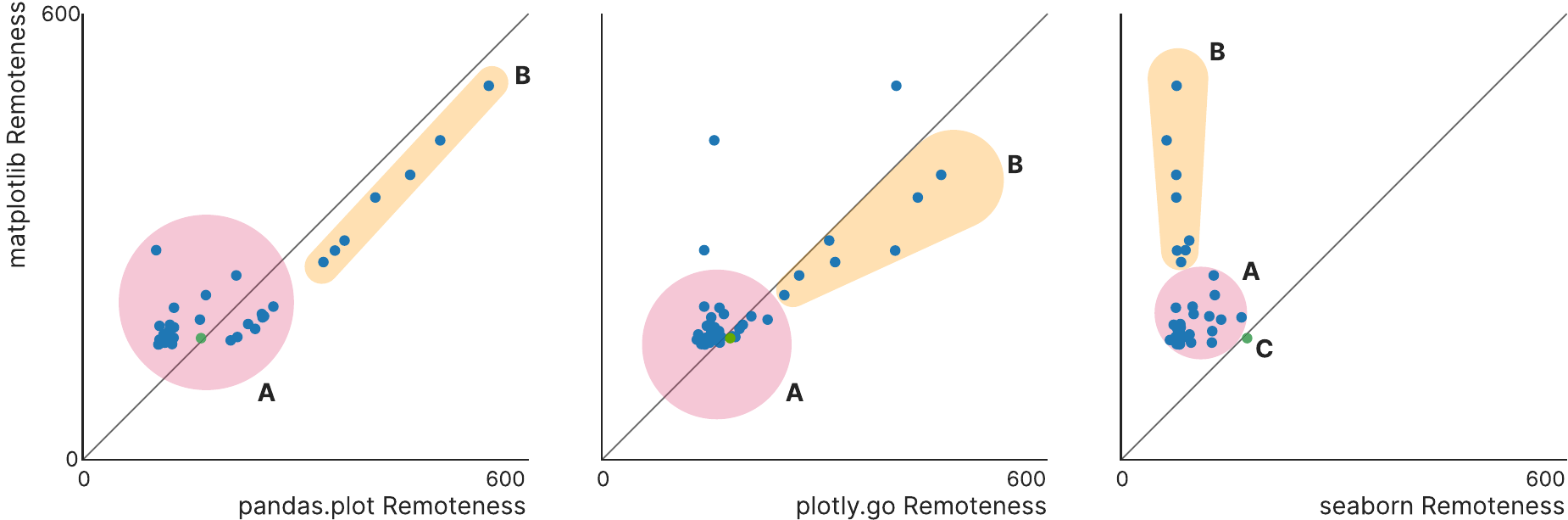}
  \caption{
    \pd{} (left plot, horizontal axis) and \go{} (middle) display similar remoteness distributions compared to \mpl{} (vertical axes), with a small core (A) of less-remote specifications and a tail (B) of more-remote ones. \sns{} (right) has lower remoteness for all but one spec (C) and a lower sprawl as a result.
  }
  \label{fig:matplotlib_remoteness}
  \vspace{-2em}
\end{figure}

\parahead{High-level observations}
We begin by considering the high-level relationships among the notations revealed by the metrics.
\figref{fig:remote-distro} highlights the presence of long tails in the remoteness distributions and the large number of tokens used only once in certain notations,  suggests that any conclusions we draw may be sensitive to outliers (and hence to our selection of examples and our implementation choices).
We used bootstrapping to understand the relationship between our metrics while taking into account this uncertainty: \figref{fig:clusters} shows the contours of the resulting distributions.

We relate these high-level patterns to Wongsuphasawat's ranking of programmatic approaches~\cite{2020EncodableConfigurableGrammar, 2020WorldVisLib}---which span from \emph{graphics library}, \emph{low-level composition}, \emph{grammars}, \emph{high-level composition}, to \emph{chart templates} (ordered by level of abstraction).
Our measurements of this gallery in \figref{fig:clusters} show one major outlying cluster consisting of those with high sprawl and large vocabularies (\eg\ \mpl{}, \pd{}, \go{}).
These share features with the \emph{high-level composition} approach, based on specifying how multiple \textit{series} should be drawn.
Next, those with small sprawl or vocabularies (\vgl{}, \alt{}, \ggp{}) resemble the \emph{grammar} approach.
The remaining notations (\sns{}, \px{}, and \sbo{}) most closely resemble the low effort and expressiveness \emph{chart templates}.
Our results contrast with Wongsuphasawat's ranking (which assumed decreasing effort and expressiveness with increasing abstraction)
as all notations were equally-able to express our gallery.
Notably, we did not observe a fundamental difference between DSL-based notations (\vgl{}) and those based in general-purpose languages.

Based on Iverson's arguments about \dimension{economy}, we expected to see an inverse relationship between sprawl and vocabulary size.
However, the results complicate the picture: larger vocabularies seem associated with \emph{higher, not lower} sprawls, and we see a range of median spec lengths associated with large vocabularies.
Part of the increase in vocabulary size for \mpl{}, \pd{}, and \go{} consists primarily of tokens not explicitly used for visualization but for general-purpose imperative Python tokens.
The other notations leverage more declarative constructs, which appear to require fewer unique tokens.
Next, these notations rely less upon inference of default values and more on explicit specification of behaviour, which requires more unique tokens.
Finally, the additional tokens may permit the specification of a larger range of computations beyond visualization than the smaller set of visualization-focused tokens used in the other notations.

A more subtle challenge to the expected \dimension{terseness}-\dimension{economy} tradeoff concerns the other two clusters.
The \emph{template-like} notations have larger vocabularies without having lower sprawls or spec lengths, which is explained by their heavier reliance on a general-purpose data transformation library than the \emph{grammar-like} ones.
\sns{}, \sbo{}, and \px{} have limited ability to express aggregations implicitly along with visual mapping and so must explicitly leverage Pandas for these operations.
This same effect impacts \ggp{}, which leverages dplyr for the same reason but does so less often, as dplyr appears to be less verbose than Pandas (although similar analysis of data transformation notations is important future work).
This suggests that the expected terseness benefits of enumerating chart types (as \px{} does) may well be erased by the length of the additional code required to transform data into the required shape first.

\begin{figure}[t]
  \centering
  \includegraphics[width=\columnwidth]{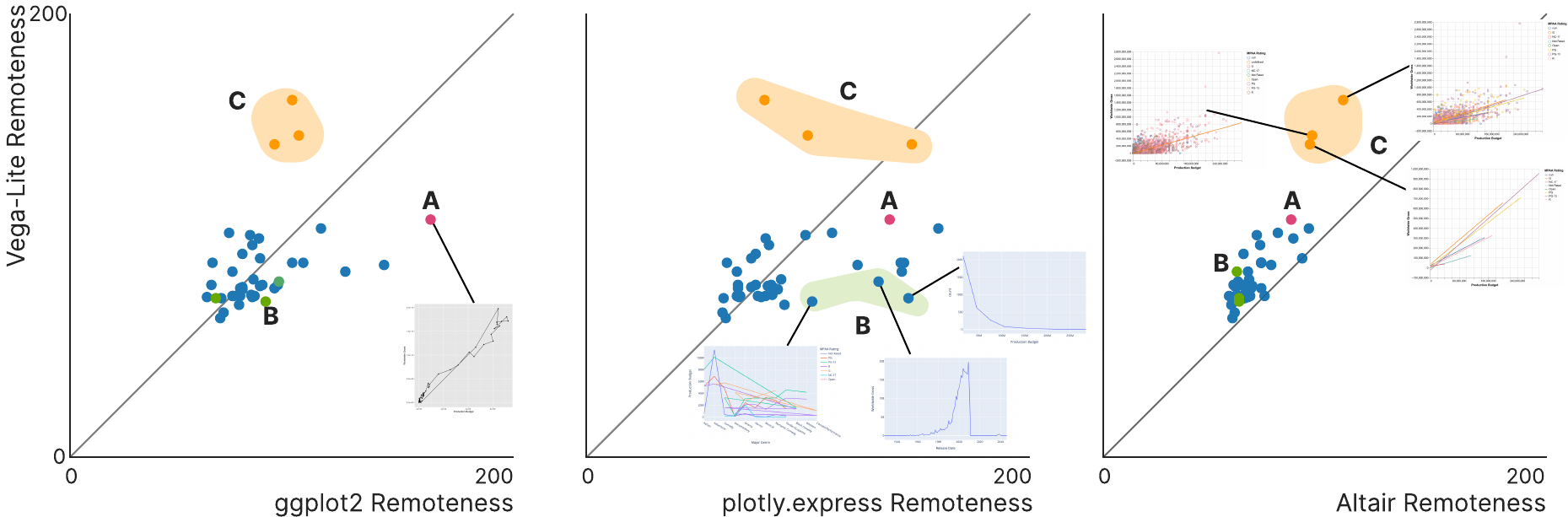}
  \caption{
    \alt{} has identical semantics and output to \vgl{}.
    Yet, \alt{} has a lower remoteness, likely caused by its ability to infer data types.
    \vgl{}, \ggp{}, and \px{} have comparable remotenesses for most specs, but diverge for connected scatterplots (A, more remote in \ggp{}), aggregated line-charts (B, more remote in \px{}), and regression lines (C, more remote in \vgl{}).
  }
  \label{fig:vega-lite_remoteness}
  \vspace{-1em}
\end{figure}

\begin{figure}[t]
  \centering
  \includegraphics[width=\columnwidth]{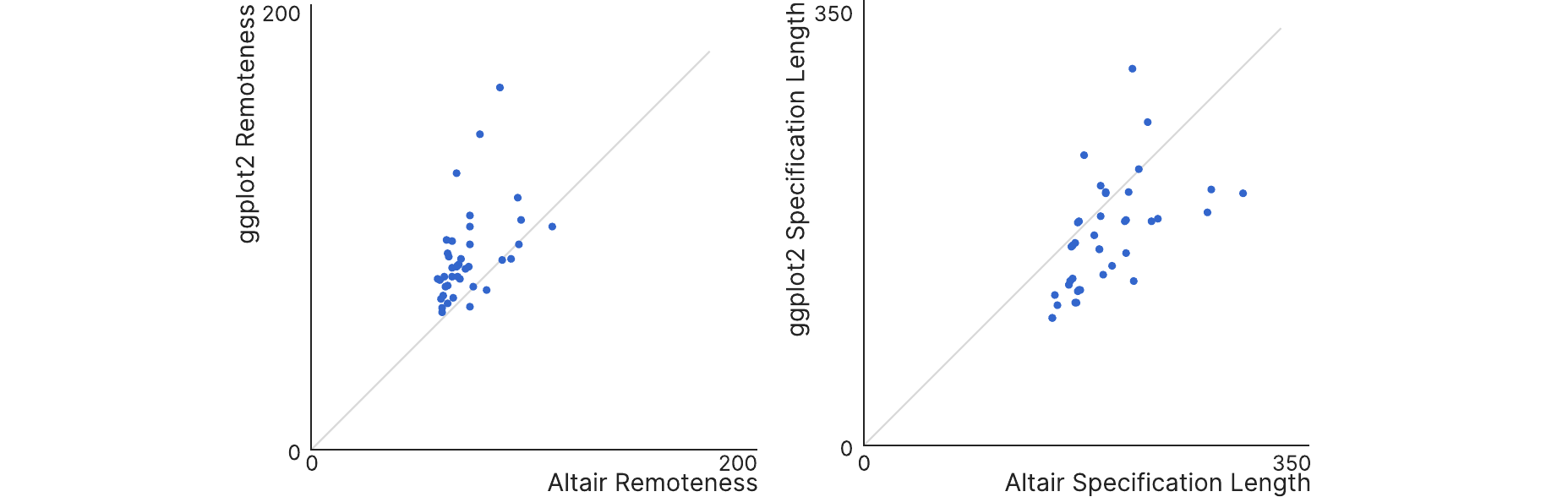}
  \caption{
    Due to their common ancestry as grammars of graphics~\cite{2005_GrammarGraphics}, \alt{} and \ggp{}
    are closely related yet have rather different sprawls and vocabulary lengths.
  }
  \label{fig:ggplot-v-altair}
  \vspace{1em}
\end{figure}

\parahead{Pairwise comparisons of notations}
While evaluation of these notations as an ensemble is useful, consideration of them in pairs also offers interesting insights.
Figs.~\ref{fig:matplotlib_remoteness}-\ref{fig:ggplot-v-altair} show several such comparisons in which
specs of similar remoteness between notations fall along the diagonal line, with patterns of diverging remoteness above or below it.

Consider the differences between \pd{} and \mpl{} shown in \figref{fig:matplotlib_remoteness}.
\pd{} leverages Pandas' \dimension{terse} but incomplete statistical graphics wrapper around certain \mpl{} functions.
As a result, some portion of our gallery could be expressed in shorter specs with lower remoteness (to the left of the diagonal) and a smaller vocabulary. In contrast, the rest could only be expressed using \mpl{}-style notation.
Although identical to the specs in \mpl{}, this latter set displays higher remoteness (to the right of the diagonal) in \pd{} because they are so different from the former set.
From this, we conclude that while providing "shortcuts" for common operations can increase the \dimension{terseness} of some parts of a notation, it can also increase its overall \dimension{viscosity}. \sns{}'s statistical graphics functionality more thoroughly covers the set of examples in our gallery, yielding smaller remoteness (and spec length) than  \mpl{} across the gallery.
This shows that a holistic approach to controlling \dimension{terseness} can also control \dimension{viscosity}, with or without a compositional, grammar-like design.

\figref{fig:ggplot-v-altair} supports this conclusion when comparing \alt{} to \ggp{}. Most \ggp{} specs are smaller than their equivalent in \alt{}, but the \alt{} specs have lower remoteness. In terms of the usability of these notations as experienced by a specification author, an initial \ggp{} spec might be quicker or easier to produce than the equivalent \alt{} one (because both ggplot2 and dplyr are highly \dimension{terse}), but iteration might be easier in \alt{} than in \ggp{} (because Altair/Vega-Lite handle most data transformations internally and implicitly).

\begin{figure}[t!]
  \centering
  \includegraphics[width=\linewidth]{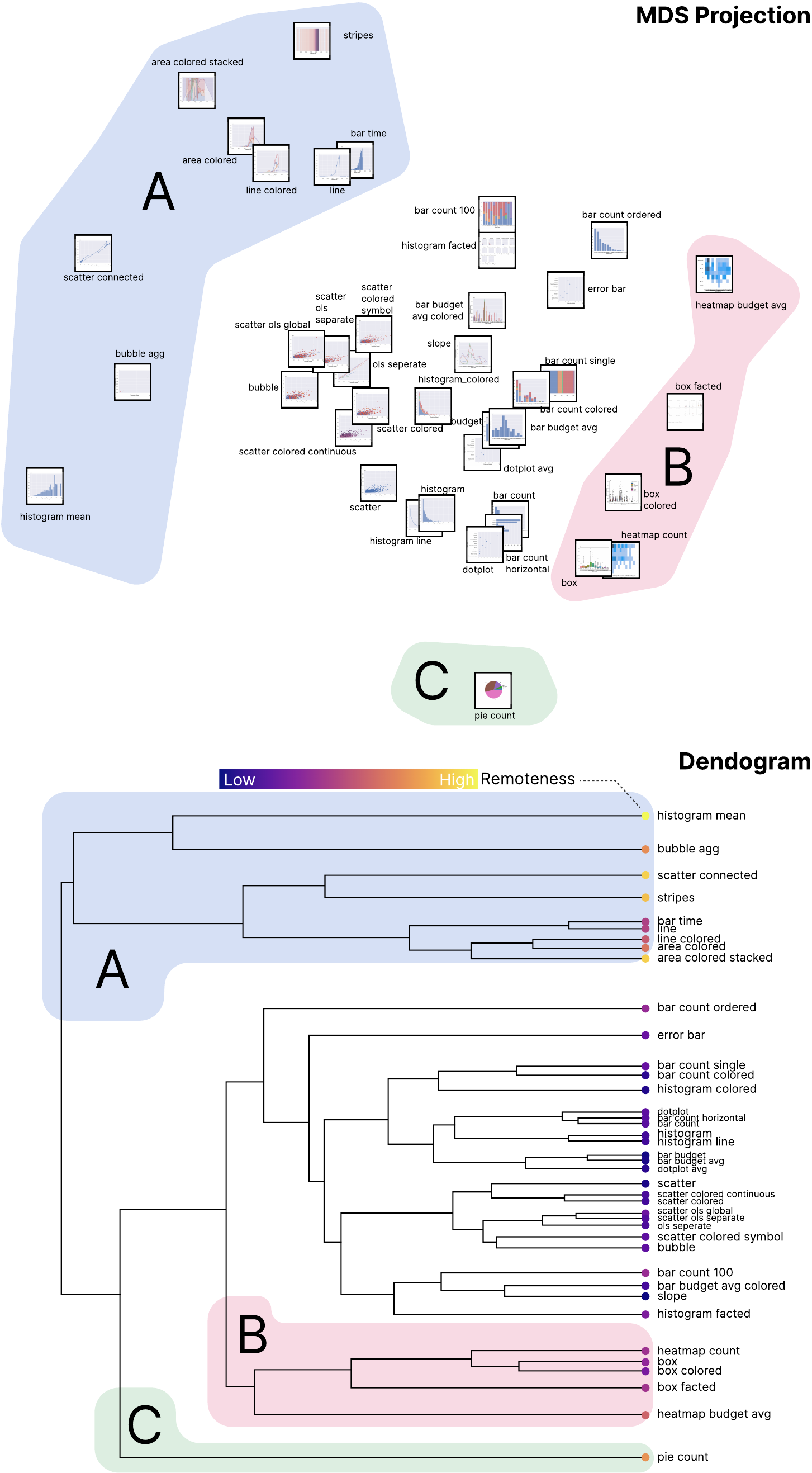}
  \caption{
    Several clusters (A, B, C) are visible for \sbo{} in an MDS embedding~\cite{mead1992review} and a dendrogram built via agglomerative clustering.
    Specs in A contain use of Pandas for data transformation, while specs in B require a fallback to \sns{}-style notation.
    C is a clear outlier as neither Seaborn API supports pie charts, prompting fallback to \pd{} notation.
  }
  \vspace{-2em}
  \label{fig:seaborn-obj}
\end{figure}

\parahead{Evaluating a single notation}
Finally, variation within a notation offers a useful perspective for analysis.
For instance, \figref{fig:seaborn-obj} shows the design space implied by our distance function for \sbo{}.
\sbo{} uses a new and partially-implemented grammar-oriented API within \sns{}, the bounds of the new API are immediately apparent, such as how \sbo{} must fall back to \sns{} in \figref{fig:seaborn-obj}B.
Finer-grained structures are also visible.
To wit, the examples seem to be clustered by chart type (scatterplots vs. bar charts) rather than by data column or metadata.

Such patterns can be compared against design goals for a notation, user intuitions about the design space, or observed or anticipated usages. For example, a notation designer may anticipate that authors will want to easily switch between certain visual forms during analysis, or may want to encourage such behaviour. Charts like \figref{fig:seaborn-obj} can be used to confirm that the representative specs are indeed clustered together.

\subsection{Expert Interviews}
\label{sec:interviews}

We gathered feedback on our approach (and its case study instantiation) by conducting  semi-structured expert interviews ($n=6$) with the authors or maintainers of ggplot2, Vega-Lite, Seaborn, Matplotlib, Altair, and Plotly.
The interviews consisted of predefined open-ended questions that sought to elicit attitudes about notation design (see the osf for instrument), followed by a discussion of our case study.
Interviews lasted about an hour.
They were conducted remotely and  automatically transcribed, with consent.
Experts were invited to participate if it was possible to identify them as playing key roles in a prominent charting library.
Participation was not compensated.
Interviews were analyzed by the first author, the results of which were refined via discussion with the rest of the project team.\footnote{
  Participants were randomly assigned ids (\eg\ \pxx{X}) and are quoted \qt{like so}.
}

\parahead{On the Approach}
All participants offered positive attitudes towards our metrics-enhanced approach to analyzing notations exemplified in \secref{sec:analysis}.
For instance, \pxa{} expressed their enthusiasm for the approach noting that \qt{anyone who's designing an API wants to make it easy and this gives information about what's easy and where exactly the deeper parts of the API have to be brought in.}
Similarly, \pxb{} noted that the types of questions surfaced by this approach are \qt{absolutely something that I think about,} going on to observe that this approach \qt{feels like it's formalizing intuitions that I have.}
While difficult to validate a method generally, \pxe{} observed that \qt{I think this  captures my mental model in a quantitative way} and went on to note that \qt{the fact that this confirms my priors gives me faith that this is actually measuring what you say it's measuring.} \pxc{} found the approach to be\qt{an interesting lens to think about things,} but \figref{fig:clusters} was \qt{not super-compelling} in that it did not challenge his expectations as it did ours, suggesting that the metrics successfully create grounds for comparing mental models among experts.
Further, participants felt that the analysis captured something important about their systems and goals. For instance, \pxc{}, \pxb{}, and \pxd{} identified the \metricName{remoteness} vs. \metricName{vocabulary size} graph (\figref{fig:clusters} lower-right) as reflecting concerns that occurred in their notational design experiences. \pxb{} and \pxd{} mapped their normative stance onto the metrics, with \pxd{} noting, \qt{Bottom-left is good, number of tokens is low, less things to learn.}

Beyond their overall reactions, participants came up with some additional ways to use our metrics-based approach.
For example, \pxa{} wanted to use the multi-notation gallery to \qt{learn from outliers} and figure out which \qt{things that are easier in other} notations.
Others observed that these metrics could facilitate conversations within or across team boundaries or with students.
\qt{It's a nice prompt for thinking about some of these design principles that are not even written down anywhere, don't consciously think about them...this helps advance that conversation} (\pxd{}) and specifically around novices or new team members: \qt{If there's a way to get these shared assumptions into new people's heads faster that's great} (\pxe{}). \pxf{} felt it \qt{could be useful for me to motivate our students. You know, why do we pick a high level language versus a low level one? Just that separation seems really clear.}
Such desired usages underscore the value of our method's explicit surfacing of measures as a way to extend shared mental models and re-see familiar notations in new lights.

Turning to our choice of metrics, experts were similarly enthusiastic.
Interviewees used a variety of spatial language to describe their reasoning about API design, noting the desirability of \qt{being able to move to near-by plots, which clearly does imply some distance} (\pxc{}) and \qt{exploring the neighborhood} (\pxd{}) around a chart.
This suggests that measures like our \metricName{sprawl}
(that describe the notational connectedness)
are valid measures of its usability.
Similarly, some experts seemed to use specification length as a proxy for complexity, therein supporting our choice of metric.
For instance, \pxd{} noted that he tried to design his notation such that
\qt{common things that seem simple should be simple and uncommon things complex.}
This suggests that the family of metrics we selected aligned with participants' concerns. That said, several participants also evoked concerns related to \dimension{progressive evaluation}.
\pxc{} asserted \qt{you want to be able to navigate from one part of the space to another...in a sequence of small steps}, while \pxd{} noted that they aspired to enable
\qt{Incremental, atomic change equals new chart.}

Participants also had reservations about our analysis method.
Every expert had suggestions for additional notations to be included in the analysis, and for the inclusion of interaction, styling, or layout features in the current ones.
Some participants echoed \pxd{}'s \qt{So many other concerns to think about when choosing a language that this doesn't capture,} such as documentation, \dimension{consistency}, and \dimension{progressive evaluation}.
Further, \pxc{} noted that this approach is \qt{more useful at the overarching scale, not driving decisions.} and \pxb{} observed that \qt{This doesn't help you find the breaking point,} \ie{} the boundaries of the subspace beyond which a high-level notation no longer provides \dimension{terseness} beyond its lower-level constituent parts. \pxe{} echoed this concern .

\parahead{On Notation Design}
Participants expressed various views about effective notation design, however these views were typically not based on formal foundations.
Certain projects have made some of their principles explicit~\cite{tidyverseContrib, vlContrib}, but most experts indicated that they relied strongly on intuition or conversation with trusted users to make decisions.
Further, none of them referred to any evaluation frameworks as guiding their designs (such as CDN~\cite{green1989cognitive}), underscoring the potential utility of making such measurements easily usable.

Participants were cognizant of their ability to shape users' experience of visualization and how they perform analysis. For instance, \pxe{} noted \qt{APIs encode a set of assumptions, be careful about the assumptions you're making, what will you prevent.}
Some experts saw their work as \qt{an opportunity to use these languages to teach about what visualizations are}(\pxd{}) or to guide users \qt{You definitely don't want everything to be equidistant because you want to help guide people's choices}(\pxc{}).
In some cases, the experts evoked the notion of an exogenous design space, which the notation makes easier or harder to reach, rather than the notation imposing those distances directly. \pxc{} argued that there should be \qt{bits that are easier to get to and others that you have to climb up a hill} to reach.

Next, participants' approaches to \dimension{economy} were generally in agreement, even if they diverged on the importance of \dimension{terseness}.
For instance, \pxa{} described their approach as \qt{not being too concerned with how terse the representation is but how descriptive,} with \pxe{} agreeing. This contrasts with  \pxb{}'s observation that \qt{I definitely aim for terseness} to support rapid iteration, echoed by \pxf{} and \pxd{}.
\pxc{} took a position between these, observing that \qt{You want to strive to keep the core vocabulary as small as possible} while balancing \qt{between breadth of vocabulary and the terseness.}.He summarized his position: \qt{If you keep adding functions, the cost is someone learning the API and if it just dissolves into 100s of special cases then you've lost the benefit of having a compositional API}.
This key tension in visualization notation design is readily exposed by our approach and would seem to suggest its potential value for future analyses.

Finally, experts espoused concordant views around the point-in-time, work-in-progress nature of the notations defined by their systems and the importance of \dimension{consistency}.
Every interview also touched on the importance of backward-compatibility. This was sometimes framed in terms of consistency with the existing API, even if it embodies what are now considered mistakes: \qt{You're going to make mistakes... Are you going to choose to be consistent with the mistake you made previously, or are you going to choose to be consistent with the underlying principle?} (\pxc{}).
\pxa{} and \pxc{} also spoke of forward-compatibility---for instance, \pxa{} observed that \qt{I try to think through not just the actual feature but what are the possible extensions of the feature that might come up in the future.}
Every expert expected to continue to evolve their notation, and some appreciated seeing comparisons between old and new notations, \eg{} \sns{} and \sbo{}.
Further, this suggests that forming metrics to measure \dimension{consistency} would be especially useful.
\section{Conclusion and Future Work}

In this work, we demonstrated how metrics computed from a multi-notation gallery can provide fertile grounds on which to analyze and compare visualization notations.
In doing so, we demonstrated how designers of such notations can use this style of reading to re-see the context of their work, findings which we evaluated through interviews with domain experts.
We now reflect on each aspect of our work, with a focus on limitations and opportunities for future work.

\parahead{Approach} Our metrics-based approach to reading visualization notations offers new tools with which to evaluate and compare visualization notations.
Using this approach, authors of new notations might evaluate whether they have successfully achieved their design goals and navigated design tradeoffs, or they might point to quantitative \metricName{remoteness} comparisons when discussing relative notational expressiveness.

Our approach is based on a metrics-enhanced close reading~\cite{bares20Close}, known as a near-by reading~\cite{mcnutt2020supporting}.
In line with this, we stress again that the computed metrics (like \metricName{sprawl} or \metricName{vocabulary size}) are not normative evaluation measures to be minimized at all costs.
Rather our intention is that the data from which they are derived can help inform more nuanced evaluations than are common in research or practice today.
We suggest that this approach enables users without a background in cognitive usability evaluation to apply those ideas by re-framing them in a more familiar data analysis-style context (as exemplified by NotaScope).
The goal of this paper was to document and demonstrate our method.
In future work, we intend to compare it with other methods---such as unassisted CDN~\cite{green1989cognitive}, critical reflections~\cite{satyanarayan2019critical}, or algebraic analyses~\cite{mcnutt2021table, pu2020probabilistic}.
A limitation of our work is that it treats computable metrics as proxies for  usability-oriented qualities~\cite{iverson1979notation, green1989cognitive}.
Future work should consider the relationships between our metrics and such quantities empirically.
For instance, the relationship between compression distance and the cognitive effort required to navigate the design space should be explored.
In addition, notation usability may be impacted by factors not contained within the text of the specs. These include the intuitiveness of the terms used, which are likely guided by cultural conceptions of data~\cite{rakotondravony2022probablement}.
It is also likely guided by sociotechnical factors~\cite{de2018library}, such as popularity, tool support, or documentation quality---which was also noted by \pxd{}.
Further, analyses using our method may be biased by our choice, design,  implementation of example metrics, as well as the selection and construction of gallery examples.
However, such issues are also endemic to single-notation gallery-based evaluation.
Our approach improves on current practice by encouraging more structured comparisons, but it does not address all issues.

Finally, there has been an explosion of interest in the visualization community~\cite{mcnutt2022noGrammar, pu2021special} around creating systematic descriptions of various practices through grammars or DSLs (which we capture via the more general notion of \emph{notation}).
While the power of such designs should not be understated, without a way to systematically examine new notations, the success of such efforts cannot be reliably evaluated~\cite{pu23Grammar}.
This work takes steps towards providing structured means for notational comparison that does not rely on costly user studies or having substantial system-building experience~\cite{satyanarayan2019critical}.
Future work might explore standardized task- or domain-oriented benchmarks galleries (analogous to benchmark suites from other fields~\cite{tpc2010tpc}), providing even more structure.
In addition, future work might extend our approach by introducing an interrogatory antagonist or creating more structured prompts for reflection (\ala{} LitVis~\cite{wood2018design}) as a companion to the metrics.

\parahead{Multi-notation Galleries} We introduce the idea of task- or domain-focused multi-notation galleries (which extend single-notation galleries) as a way to make systematic comparisons of notations possible.
Our metrics augment this process by helping to identify patterns and outliers, however even without metrics, multi-notation galleries can help examiners compare notations (\ala{} close reading).
While able to answer many questions, multi-notation galleries can not bound the \dimension{expressive power} of a notation.
That is, galleries cannot demonstrate that a given chart is impossible in a particular notation.
For instance, to the surprise of its designers, an earlier version of \vgl{} without \code{arc} marks was shown to be able express pie charts~\cite{cheatingPie} through its geographic projection system.
We suggest then that \dimension{expressive power} might be recast as an arbitrary bound on a metric like \metricName{specification length} or \metricName{remoteness} instead: what our experts referred to as the notational ``breaking point.''
For instance, \pxf{} observed that he avoids showing novices certain valid specifications because they are too complex (i.e. too long) or differ too much from other semantically similar specs (i.e. too remote).
Future work should explore how such extrema might be automatically exposed to the designer.
Our galleries have a number of limitations.
Each notation must produce all of the examples, which causes them to be limited to the  \emph{greatest common feature set} among the notations, although, as noted above, this set can be quite large.
However, the existence of a common problem domain (\eg{} charting) suggests that notations are designed to fit a set of tasks that users realistically seek to accomplish. If a notation possesses features that are so unlike others as to be incomparable, then the user may be pursuing a more nuanced task (\eg{} ECharts~\cite{li2018echarts} streaming data).
Requiring the gallery to be centered on a single dataset may bias the  analyses based on the types of examples producible with that dataset, however given that data tends to flow through visualizations rather than be expressed as part of their notation, we believe this effect is limited.
Our use of metrics over galleries measures notational samples rather than the notation itself.
Direct measurement would be preferable, but it is impeded by most notations lacking formal definitions. Approaches like grammar-ware~\cite{klint2005toward} may facilitate higher order comparison, however the lack of adoption of this style of work in the last 20 years suggests that this is unlikely.
The labor intensive process of producing galleries might be reduced by employing a generative AI-based assistant~\cite{vaswani2017attention}, although that invites the tradeoffs endemic to such systems~\cite{di2023doom}.

\parahead{Case study}
Our case study, and its evaluation, demonstrated that our metrics-based approach can produce findings that experts in statistical graphics notation design find interesting.
As these experts have necessarily done the closest reading of their libraries, we suggest their finding value in our approach indicates that our metrics provide useful purchase for analysis.
While not able to draw conclusions about matters like which notation is best, this study provided empirical evidence for various patterns latent to this family of notations---such as the difference between \emph{High-Level Composition} and \emph{Visualization Grammar}-style approaches~\cite{2020EncodableConfigurableGrammar}.
Further, the distances and token-counts reveal other types of variation between notations, such as in their \dimension{viscosity}, \dimension{terseness}, and \dimension{economy}.
In future work, we intend to validate our findings by replicating our study with another gallery for a different dataset.

The selection of notations considered in our case study was motivated by the twin criteria of trying to select the most widely-used libraries and those for which we could solicit evaluation by experts, the resolution of which may have biased our results.
This study might be usefully extended to include other common statistical graphics notations (and experts therein) such as D3\cite{bostock2011d3}, base-R, Bokeh~\cite{bokeh}, Holoviz~\cite{yang2022holoviz}, AntV~\cite{antvSpec}, or ECharts~\cite{li2018echarts}.
Other visualization domains could be usefully considered---\eg{} thematic mapping, interaction, or animation---as might those outside of it---\eg{} diagrams~\cite{text2diagram}, data manipulation (SQL vs. dplyr vs. Pandas), or ML models (Torch vs. Tensorflow).

While the interviewed experts were enthusiastic about our approach and analysis, \pxc{} and \pxd{} felt that our case study (rooted in NotaScope) was too high-level for individual design decisions.
However, the metrics-based approach could be applied on a much finer-grained level: instead of just quantifying the distance between two specifications, NotaScope could be extended to automatically enumerate and display plausible step-wise paths through the design space.
The presence of multiple paths with small steps would speak to  \dimension{progressive evaluation}, which is an important (\pxc{}, \pxd{}) design goal that can be hard for designers to reason about.
Our study design limits these results,
which may have yielded different findings with more users or if we had elicited structured Likert-style ratings. In future work, we intend to take such measures as we further explore our approach's usability.

We are optimistic that, based on these lessons, we can add to our suite of metrics and continue to improve our approach and tool so that they can become useful parts of the notation designer's toolkit.

\section{Supplemental Material}

Additional material related to this paper can be found at \osf{}. This part of the supplemental material includes the source code for our tool (NotaScope), the study instrument, and the specifications used to form the basis of the gallery in our case study. Further supplemental material, which can be found on PCS, contains a video walk-through of our system, and an appendix with additional written materials. Finally, a demo of our tool NotaScope is available at \href{https://app.notascope.io/}{app.notascope.io}, where the case study materials can be browsed and analyzed interactively.

\acknowledgments{%
  We acknowledge our reviewers for their thoughtful commentary, as well as the domain experts who graciously shared for their time and perspectives with us.
  This work was supported by NSERC.%
}

\bibliographystyle{abbrv-doi-hyperref}

\bibliography{template}
\clearpage{}

\appendix{}
\section{Appendix}

\begin{figure}[ht]
  \centering
  \includegraphics[width=\linewidth]{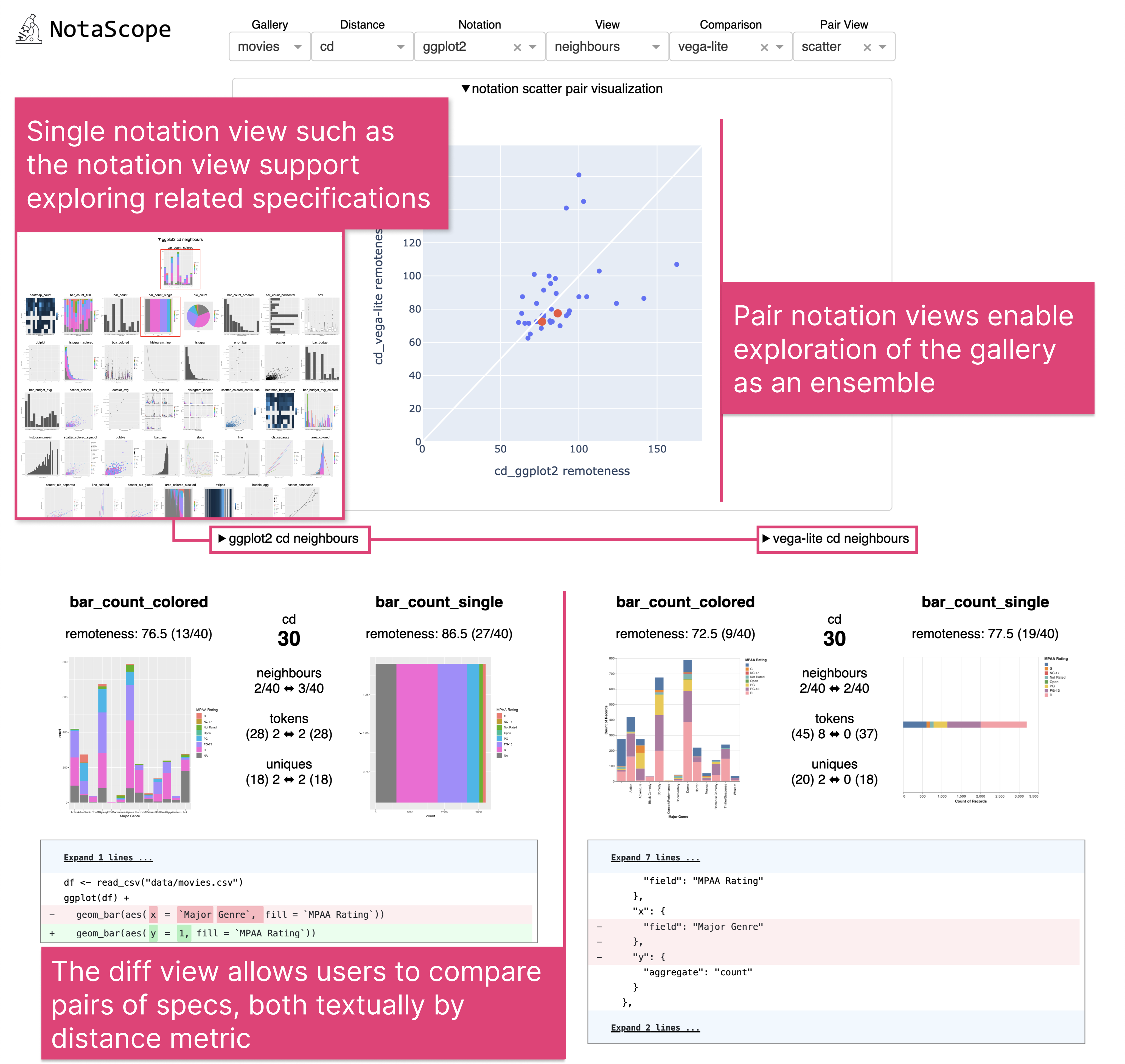}
  \caption{
  To conduct our analysis we built a tool called \href{https://app.notascope.io/}{NotaScope}, which supports a wide variety of mechanisms for comparing notations. See the video figure for a walk-through. 
  }
  \label{fig:notascope-annotated}
\end{figure}

This appendix includes material that was unable to fit into the main body of the paper.
Below we describe the notations employed in our use case. 
Then, in \figref{fig:notascope-annotated} we show an annotated screenshot of our tool, NotaScope.
Finally, in \figref{fig:vega-lite_remoteness} we exemplify a minimum spanning tree (enabled by NotaScope) of the space implied by our compression distance metric. 

\subsection{Notations}

The notations used in our case study were defined as follows:

\begin{itemize}
    \item The \ggp{} notation consists of R code constrained by having only the \code{tidyverse}~\cite{2019WelcomeTidyverse} package installed in a standard R runtime, biased towards using functions from that package rather than base R functionality wherever possible. The \code{tidyverse} package includes the ggplot2 visualization system~\cite{wickham2010layered} as well as the dplyr~\cite{dplyr} data-transformation system. \ggp{} consists of a grammar based on Wilkinson's~\cite{2005_GrammarGraphics}.
    \item The \mpl{} notation consists of Python code constrained by having only the \code{matplotlib}~\cite{hunter2007matplotlib} and \code{pandas}~\cite{mckinneyProcScipy2010} packages installed in a standard Python runtime, biased towards using Matplotlib's axis-level function API rather than the legacy \code{pyplot} API, as recommended in the documentation. This is an imperative API with functions to progressively mutate a figure. \mpl{} does not have any data-transformation capabilities, so pandas is used wherever necessary in this notation. \mpl{} is the most-downloaded Python visualization system and Pandas is the most-downloaded data-frame system used to manipulate tidy data in Python.
    \item The \pd{} notation is defined in the same way as the \mpl{} notation, except that Pandas' built-in \code{.plot()} API is used wherever possible. This API is a thin wrapper around Matplotlib to “easily create decent-looking plots”~\cite{pandasPlotDocs} We include this notation as distinct from \mpl{} due to the popularity of Pandas and to study the effects of a minor variant of a notation on the measures we have developed.
    \item The \sns{} notation consists of Python code constrained by having only the \code{seaborn}~\cite{waskom2021Seaborn} package installed in a standard Python runtime, biased towards using \sns{}'s figure-level API. Seaborn is itself a wrapper around Matplotlib and also depends on Pandas. It has some data-transformation functionality built-in and pandas' capabilities are used wherever necessary. \sns{}'s figure-level API is built around just three functions, for “distributional”, “categorical” and “relational” figures. It is the most-downloaded statistical-graphics-focused library in Python.
    \item The \sbo{} notation is defined in the same way as the \sns{} notation, except that the new, work-in-progress object-level API is used wherever possible. The object-level interface was developed to be a more consistent and extensible interface than the figure-level one. Its grammar consists of Mark, Stat, Move, and Scale objects. We include this notation to study the differences between two notations that share many design decisions.
    \item The \go{} notation consists of Python code constrained by having only the \code{plotly}~\cite{plotly} and \code{pandas} packages installed in a standard Python runtime, biased towards using Plotly's lower-level \code{graph\_objects} interface. The \code{graph\_objects} interface enables Python users to generate JSON figure descriptions to be rendered by the Plotly.js library, and includes almost no data-transformation features, so pandas' data-transformation capabilities are used. The structure of this API is oriented around the accumulation of trace objects which represent series to be drawn. Plotly is the second-most-downloaded visualization library in Python.
    \item The \px{} notation is defined in the same way as the \go{} notation except that the built-in high-level Plotly Express API is used whenever possible. Plotly Express was developed to enable the creation of terser specifications than is possible with the \go{} notation, by including some data-transformation capabilities. Wherever those capabilities are insufficient to specify an example, pandas' are used in this notation. Plotly Express is included within the \code{plotly} package and was designed to have a similar relationship to the \go{} notation as the \sns{}'s figure-level interface does to \mpl{}.
    \item The \vgl{} notation consists of JSON code constrained by having only the \code{vega-lite}~\cite{satyanarayan2016vegalite} module installed in a standard NodeJS runtime. \vgl{} is based on a highly consistent and orthogonal grammar, with built-in data-transformation capabilities.
    \item The \alt{} notation is defined by having  only the \code{altair}~\cite{vanderplas2018altair} and \code{pandas} packages installed in a standard Python runtime. Altair is a Python interface to Vega-Lite.

\end{itemize}

\onecolumn

\begin{figure}[t]
  \centering
  \includegraphics[width=0.75\linewidth]{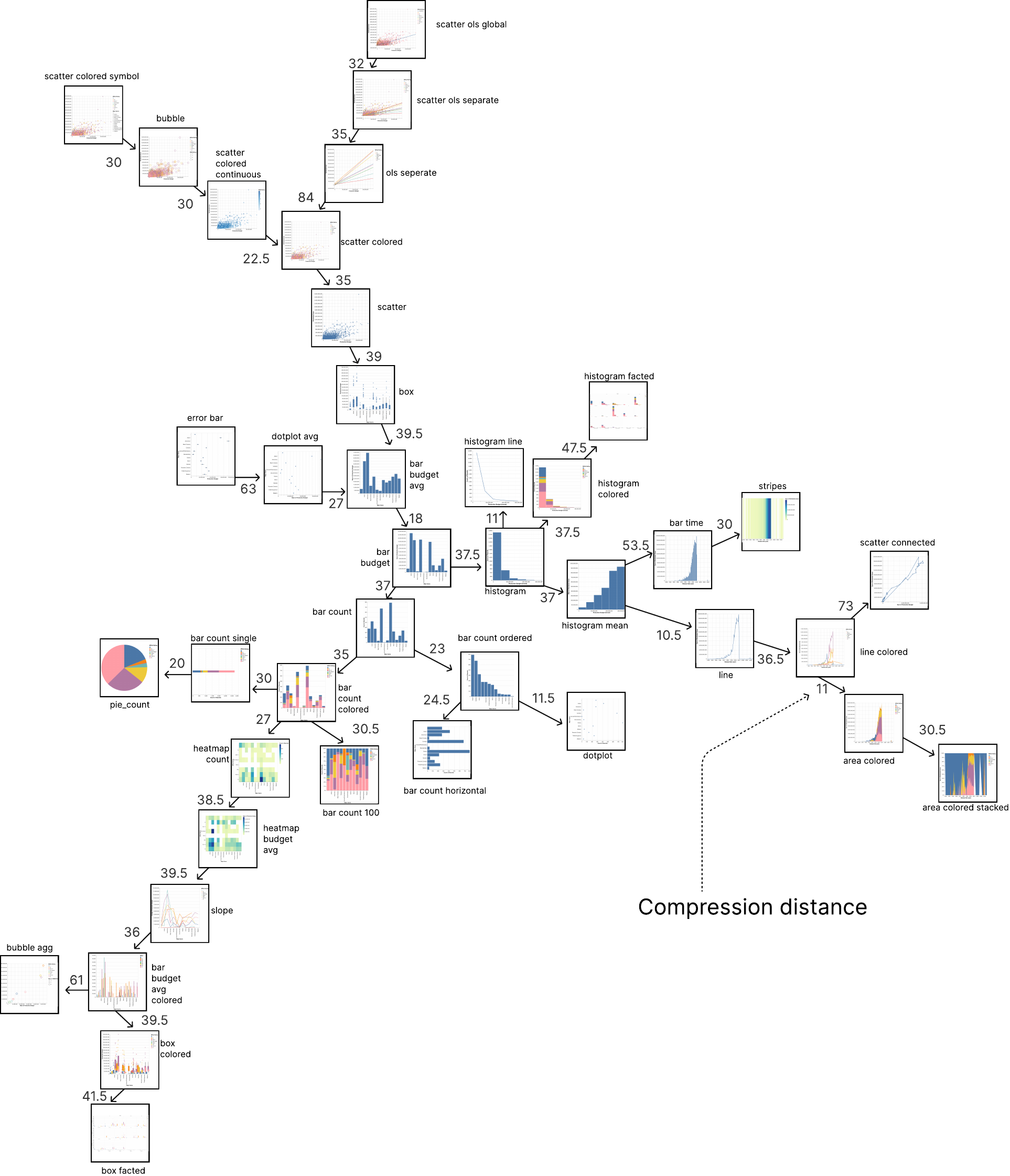}
  \caption{ 
    Notascope is capable of producing a wide range of visualizations, beyond those that we explore in the paper. For instance, here we show a minimum spanning tree for the \vgl{} specifications from our gallery using our compression distance metric. Such visualizations allow the reader to explore a path of minimum alteration through the space described by the gallery.
  }
  \label{fig:mst}
\end{figure} 
\end{document}